\documentclass[twocolumn,prb,aps,floatfix,showpacs,superscriptaddress]{revtex4}
\usepackage{graphicx}
\usepackage{dcolumn}
\usepackage{bm}

\begin{document}

\title{Dilute magnetic semiconductor and half-metal behaviors in 3\emph{d} transition-metal doped black and blue phosphorenes: a first-principles study}
\author{Weiyang Yu}
\affiliation{International Laboratory for Quantum Functional Materials of Henan, and School of Physics and Engineering, Zhengzhou University, Zhengzhou, 450001, China}
\affiliation{School of Physics and Chemistry, Henan Polytechnic University, Jiaozuo, 454000, China}
\author{Zhili Zhu}
\affiliation{International Laboratory for Quantum Functional Materials of Henan, and School of Physics and Engineering, Zhengzhou University, Zhengzhou, 450001, China}
\author{Chun-Yao Niu}
\affiliation{International Laboratory for Quantum Functional Materials of Henan, and School of Physics and Engineering, Zhengzhou University, Zhengzhou, 450001, China}
\author{Chong Li}
\affiliation{International Laboratory for Quantum Functional Materials of Henan, and School of Physics and Engineering, Zhengzhou University, Zhengzhou, 450001, China}
\author{Jun-Hyung Cho}
\email[e-mail address:]{chojh@hanyang.ac.kr}
\affiliation{Department of Physics and Research Institute for Natural Sciences, Hanyang University, 17 Haengdang-Dong, Seongdong-Ku, Seoul 133-791, Korea}
\affiliation{International Laboratory for Quantum Functional Materials of Henan, and School of Physics and Engineering, Zhengzhou University, Zhengzhou, 450001, China}
\author{Yu Jia}
\email[e-mail address:]{jiayu@zzu.edu.cn}
\affiliation{International Laboratory for Quantum Functional Materials of Henan, and School of Physics and Engineering, Zhengzhou University, Zhengzhou, 450001, China}
\date{\today}
\begin{abstract}

\textbf{Abstract}

We present first-principles density-functional calculations for the structural, electronic, and magnetic properties of substitutional 3\emph{d} transition metal (TM) impurities in two-dimensional black and blue phosphorenes. We find that the magnetic properties of such substitutional impurities can be understood in terms of a simple model based on the Hund's rule. The TM-doped black phosphorenes with Ti, V, Cr, Mn, Fe and Ni impurities show dilute magnetic semiconductor (DMS) properties while those with Sc and Co impurities show nonmagnetic properties. On the other hand, the TM-doped blue phosphorenes with V, Cr, Mn and Fe impurities show DMS properties, those with Ti and Ni impurities show half-metal properties, whereas Sc and Co doped systems show nonmagnetic properties. We identify two different regimes depending on the occupation of the hybridized electronic states of TM and phosphorous atoms: (i) bonding states are completely empty or filled for Sc- and Co-doped black and blue phosphorenes, leading to non-magnetic; (ii) non-bonding \emph{d} states are partially occupied for Ti-, V-, Cr-, Mn-, Fe- and Ni-doped black and blue phosphorenes, giving rise to large and localized spin moments. These results provide a new route for the potential applications of dilute magnetic semiconductor and half-metal in spintronic devices by employing black and blue phosphorenes.

\textbf{Keywords:} Dilute magnetic semiconductor, half-metal, transition metal doping, phosphorene

\end{abstract}

\pacs{73.22.-f, 75.50.Pp, 75.75.+a}
\maketitle

\textbf{1. Introduction}

Two-dimensional (2D) materials, graphene and silicene, are currently the subject of intense theoretical and experimental research especially for their novel electronic device applications.\cite{Geim, Castro} Graphene and silicene have demonstrated many exquisite phenomena originating from the characteristic conical dispersion and chiral behavior of their valence and conduction bands around the Fermi level.\cite{Novoselov,Katsnelson,Park1} Generally speaking, the nanostructures of graphene and silicene such as nanoribbons, nanotubes and their interconnections have opened new routes for experimental and theoretical studies in the field of nanoelectronics.\cite{Avouris} Very recently, black phosphorene, a single layer of black phosphorus (BP) was successfully fabricated through exfoliation from the bulk black phosphorus,\cite{Reich} and therefore becomes, besides graphene and silicene, another stable elemental 2D material. The black phosphorene presents some advantages superior to other previously studied 2D semiconductors because of its intriguing electronic properties, thereby drawing enormous interest from the society of materials science.\cite{Tran,Fei1,Fei2,Tranm,Zhu1,Zhu2,Yu} Recently, Li \emph{et al}\cite{Li} reported that black phosphorene could be applied to the channel of the field effect transistor (FET) device that has a high carrier mobility of $\sim$10$^{3} $cm$^{2}$/V$^.$s and an on/off ratio of $\sim$10$^{4}$ at room temperature.
As the allotrope of black phosphorene, blue phosphorene has the same stability as black phosphorene at room temperature, and its band gap is larger than black phosphorene\cite{Zhu3}.
These good electronic properties of black and blue phosphorene nanosheets can be useful for the development of future nanoelectronic devices, spintronics, and related applications.\cite{Buscema,Cai,Kulish1,Kulish2,Boukhvalov,Qin,Liu,Zheng,Dai}

For the design of practical electronic devices, defects and impurities have been employed to tune the electrical, optical, and other properties. Over the last decades the resulting of dilute magnetic semiconductors (DMS) and half-metals have achieved important developments,\cite{Dielt1,Ohno1,Ohno2,Sato,Zunger,Dielt2,Matsumoto,Coey,Zhu4,Feng} both in fundamental aspects and prospective technological applications. Indeed, it was possible to understand the underlying mechanisms of interaction between dilute magnetic impurities allowing ferromagnetic semiconductors at room temperature.\cite{Matsumoto,Coey} For prospective applications, the integration between 2D semiconductors and magnetic data storage enables the development of two-dimensional spintronics devices such as spin valve, spin-based transistors, non-volatile magnetoresistive memories and even magnetically enhanced optoelectronics devices.\cite{Zutic}

Meanwhile, in spite of the success of 2D materials such as graphene, silicene,\cite{Novoselov1} transition metal dichalcogenides (TMDCs)\cite{Novoselov2,Mak} and black phosphorene\cite{Cai,Hu2,Khan2}, there has been rare study on the dilute magnetic characters of doped 2D black phosphorene except the work of Hashmi \emph{et al}\cite{Hashmi} and Sui \emph{et al}\cite{Sui}, while half-metal properties in doped blue phosphorene have remained unexplored so far. From a technical point of view, 2D semiconductors have other superior factors that can be exploited in magnetic or spintronic devices. First, the carrier concentration can be externally controlled by voltage gating. Secondly, there is room to improve the control of the impurity concentration, for example, by employing adatoms as impurities with concentrations above the solubility limit. In practice, studies of magnetic semiconductor nanostructures with lower dimensionalities, including semiconductor nanocrystals and nanowires\cite{Dielt2,Kittilstved,Rao,Park2} doped with transition metals (TM), demonstrated that the confinement effect and the improved control of magnetic dopants can be used to increase the Curie temperature.\cite{Seixas,Ramasubramaniam}

In this work, we focus on substitutional 3\emph{d} TM impurities (from Sc to Ni) in black and blue phosphorenes to investigate their dilute magnetic characters and half-metal properties. Using first-principles density functional theory (DFT) calculations, we study the structural, electronic, and magnetic properties of substitutional 3\emph{d} TM impurities in black and blue phosphorenes. One of our key results is that the electronic and magnetic properties of these substitutional impurities can be understood by a simple model based on the hybridization between the TM \emph{d} orbitals and the defect (i.e., phosphorous vacancy) levels. This model together with the calculated band structure provides an explanation for the non-trivial behaviors of the binding energy and the spin moments for all the systems considered. Concisely, we distinguish two different regimes that depend on the electron filling of TM-phosphorous hybridized levels: (i) completely unoccupied (occupied) bonding states for Sc (Co) lead to non-magnetic; (ii) partially occupied non-bonding \emph{d} shell for Ti, V, Cr, Mn, Fe and Ni give rise to large and localized spin moments.

This paper is organized as follows. After a brief description of the computational details in section 2, we present the geometry structures, binding energies and magnetic properties of all the substitutional TM impurities studied in section 3.  We also present the general ideas behind our model of the metal-phosphorus hybridization in the considered systems. In section 4, the electronic structure of the unreconstructed D$_{3h}$ phosphorus vacancy in pristine phosphorene, along with the electronic structure of the different groups of impurities are presented. Finally, we give a summary with some general conclusions.

\begin{figure}[t]
\includegraphics[width=7.5cm]{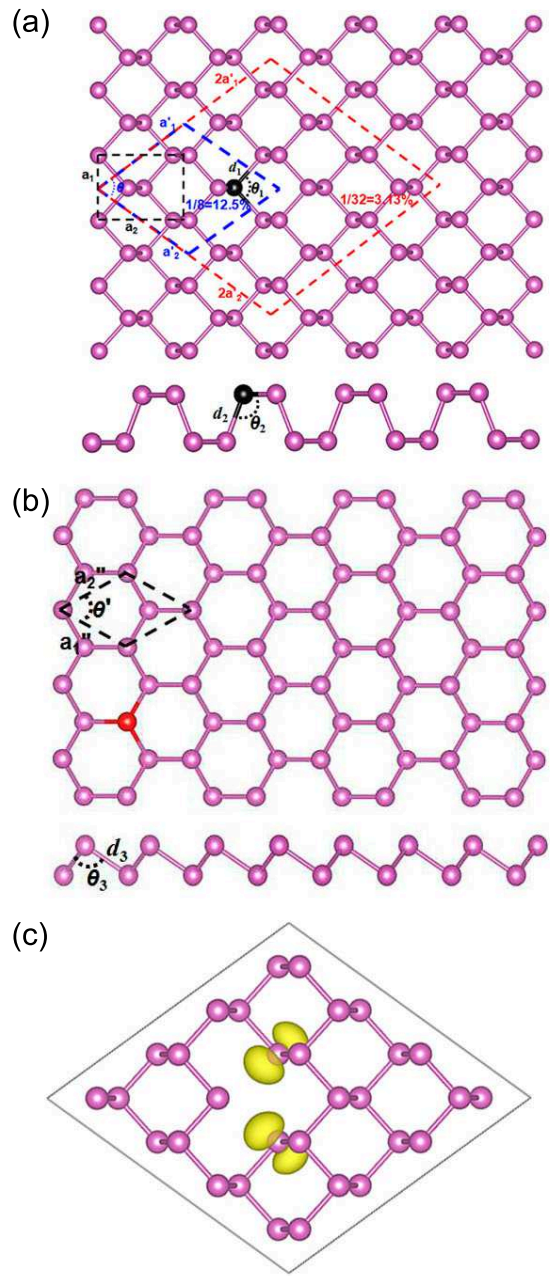}
\caption{(a) Top and side views of a diamond-like 2$\times$2 supercell of black phosphorene used in the present calculations. (b) Top and side views of blue phosphorene. (c) Spin density of black phosphorene containing a vacancy with an isosurface of 0.005 e/\AA$^3$.}
\end{figure}

\begin{figure*}[htb]
\includegraphics[width=15cm]{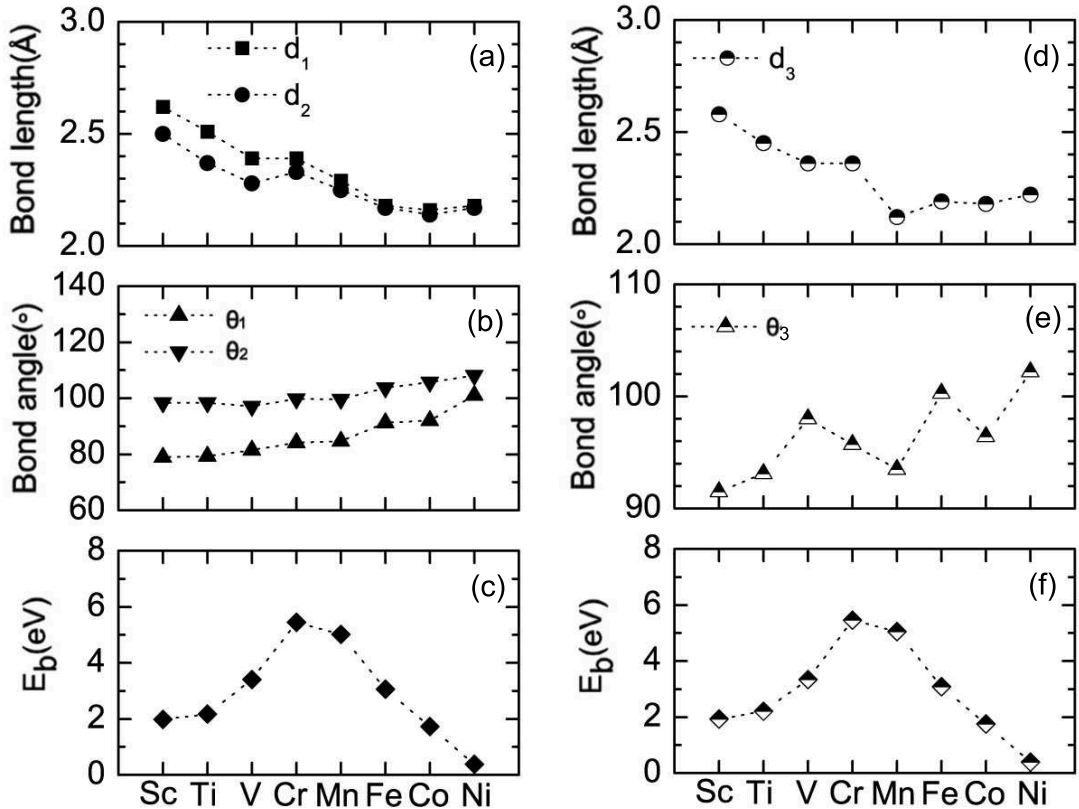}
\caption{(a), (b) Structural parameters and binding energies (c) of the substitutional TM-doped black phosphorenes. The corresponding ones of the substitutional TM-doped blue phosphorenes are given in (d), (e), and (f).}
\end{figure*}

\textbf{2. Computational Details}

The present DFT calculations were performed using the Vienna \emph{ab initio} Simulation Package (VASP) code with a plane-wave basis set.\cite{Kresse1,Kresse2} Projector augmented wave (PAW) potentials\cite{Blochl} were used to describe the core electrons and the generalized gradient approximation (GGA) of Perdew, Burke and Ernzernhof (PBE)\cite{Perdew} was adopted for exchange-correlation energy. To examine the reliability of the PBE method on the magnetic properties of the black and blue phosphorenes containing 3\emph{d} TM impurity atoms, we also considered the effect of the on-site Coulomb interaction \emph{U} on the magnetic property within the PBE + \emph{U} method.\cite{Dudarev} A kinetic energy cutoff of the plane-wave basis set was used to be 500 eV and for the structural optimization, convergence of Hellmann-Feynman residual forces less than 0.01 eV/{\AA} per atom was achieved. Because the convergence with respect to the number of $k$-points was especially critical to obtain accurate results for the spin moment in the systems studied, we used an adequate number of $k$-points for all the different supercell sizes, equivalent to 9$\times$9$\times$1 Monkhorst-Pack \cite{Monkhosrt} sampling. The Fermi level was smeared by the Gaussian method with a width of 0.05 eV. Most of our results were obtained using 2$\times$2/4$\times$4 crystallographic symmetrical supercells of black and blue phosphorenes, with a doping concentration of  3.13\%, as shown in Fig. 1(a) and (b). We also checked the results by performing calculations using larger supercells up to 4$\times$4/6$\times$6 [concentration of 1.56(1.39)\%] for several elements. In order to avoid spurious interactions between periodic images of the defective phosphorene layer, a vacuum spacing perpendicular to the plane was employed to be larger than $\sim$15 {\AA}.

\begin{figure*}[htb]
\includegraphics[width=17cm]{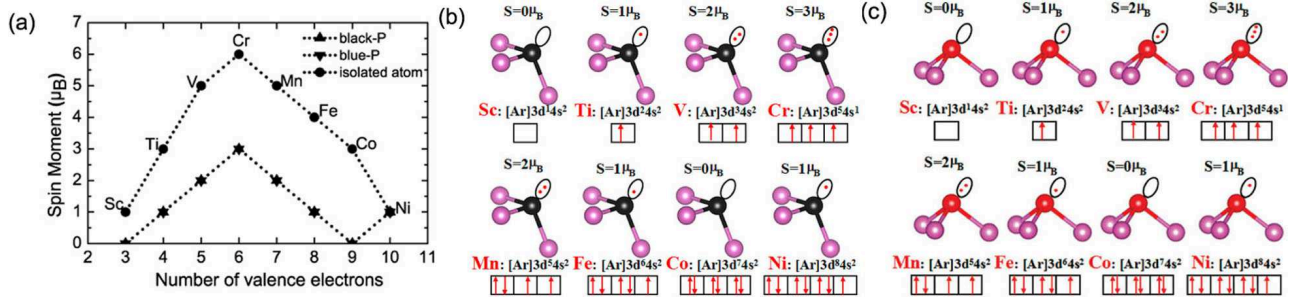}
\caption{(a) Spin moments of the isolated TMs and their substitutions in black and blue phosphorenes as a function of the number of valence electrons (Slater-Pauling-type plot). Schematic diagram of spin moment in doped black phosphorene (b) and blue phosphorene (c) in terms of Hund's rule, respectively.}
\end{figure*}

\begin{table*}
\caption{Electronic charges of each atomic species in the TM-doped black and blue phosphorenes, obtained using Bader charge analysis. The positive (negative) sign represents the gained (lost) electrons.}
\begin{tabular}{p{1.2cm}p{2.0cm}p{1.0cm}p{1.0cm}p{1.0cm}p{1.0cm}p{1.0cm}p{1.0cm}p{1.0cm}p{1.0cm}}
\hline
\hline
        &Atoms     &   Sc    &    Ti   &    V    &    Cr   &   Mn    &   Fe    &   Co    &   Ni  \\
\hline
        &TM        &  -1.54  &  -1.23  &  -0.97  &  -0.87  &  -0.61  &  -0.37  &  -0.19  &  -0.16 \\
black-P &Nearest-P &  +0.97  &  +0.78  &  +0.59  &  +0.59  &  +0.33  &  +0.12  &  0.00   &  +0.07  \\
\hline
        &TM        &  -1.55  &  -1.27  &  -1.05  &  -0.88  &  -0.57  &  -0.40  &  -0.21  &  -0.20 \\
blue-P  &Nearest-P &  +0.32  &  +0.27  &  +0.26  &  +0.20  &  +0.12  &  +0.07  &  0.00   &  +0.02  \\
\hline
\hline
\end{tabular}
\end{table*}

\textbf{3. Structural, energetic, and magnetic properties of TM doped in black and blue phosphorenes}

In this section we provide our results for the geometries, binding energies, and spin moments of substitutional TMs in black and blue phosphorenes.

\textbf{3.1 Geometrical parameters and binding energies}

The typical structure of the systems studied in this paper is presented in Fig. 1. Fig. 1(a) shows the diamond-like 2$\times$2 supercell structure of monolayer black phosphorene, with a doping concentration of 3.13\%. In view of symmetry and doping isotropy, the diamond-shape unit cell was employed instead of the rectangle unit cell. As shown in Fig. 1(a), the optimized lattice constants are $a_1$=3.310{\AA}, $a_2$=4.589{\AA}, and $a_1'$=$a_2'$=($a_1^{2}$+$a_2^{2}$)$^{1/2}$=5.658{\AA}, and the angle between the basis vectors $a_1'$ and $a_2'$ is 71.89$^\circ$. These values are consistent with experiment\cite{Crichton} and other theoretical calculations\cite{Qiao}.
Fig. 1(b) displays the optimized structure of the blue phosphorene with lattice constants $a_1''$=$a_2''$=3.330{\AA}, and their angle $\theta^{'}$=60$^\circ$, which are in good agreement with previous DFT calculations\cite{Zhu3}.

We begin to study a pure black phosphorene with a monovacancy. Fig. 1(c) shows the spin density of black phosphorene with a monovacancy. Similar to the results reported by Ma \emph{et al}\cite{Ma}, the phosphorus atoms around the vacancy undergo a Jahn-Teller distortion, and two of the phosphorus atoms close to the vacancy site move towards each other to form a P-P distance of 1.832{\AA}, which is 0.408{\AA} smaller than that of  the intrinsic phosphorene. The ground state of the system has a magnetic moment of 1.00 $\mu_B$/unit cell, most of which is concentrated at the two P atoms with the unsaturated bonds, as seen in Fig. 1(c).

The structural parameters and energetic properties of the substitutional TMs in black and blue phosphorenes are shown in Fig. 2. For TM-doped black phosphorene, the bond angles $\theta_{1}$ and $\theta_{2}$ monotonically increase from Sc to Ni [see Fig. 2(b)]. Meanwhile, the bond lengths $d_1$ and $d_2$ decrease for Sc-, Ti-, and V-doped systems and increase for Cr-doped system, and then decrease again for Mn-, Fe-, and Co-doped systems, and then increase for Ni-doped system [see Fig. 2(a)]. These behaviors of the bond lengths and bond angles reflect the size of the TM atoms. As for TM-doped blue phosphorene, the bond length $d_3$ decreases from Sc to Mn and then increases from Mn to Ni [see Fig. 2(d)], while the band angle $\theta_{3}$ shows an oscillating behavior [see Fig. 2(e)].

Figure 2(c) shows the calculated binding energies ($E_b$) of the TM-doped black phosphorenes, where $E_b$ is defined as -($E_{total}$-$E_{phosphorene}$-$E_{atom}$). Here, $E_{total}$ is the energy of the whole configuration, $E_{phosphorene}$ is the energy of the phosphorene with a vacancy and $E_{atom}$ represents the energy of an isolated dopant atom. We find a continuous increase of the binding energy from Sc to Cr, and then decrease from Mn to Ni, and the binding energies for the considered TMs are in the range of 0.375-5.466 eV. Interestingly, Cr-doped system has the maximum binding energy. This peculiar behavior is related with the interplay between the energy down-shift and the compression of the 3\emph{d} shell of the TM as the atomic number increases. Although this explanation will be more clear when the metal-phosphorus hybridization levels are discussed below, we note that the behavior of the binding energies of the substitutional 3\emph{d} TM arises from two competing effects:

(i) From Sc to Cr, the decrease of $d_1$ and $d_2$ reflects an increase in the bonding strength between the TM and phosphorous atoms, and
(ii) From Mn to Ni, as the 3\emph{d} shell is occupied, its hybridization with the phosphorous vacancy states is weakened to decrease the binding energy.

It is noticeable that this trend of the energetics for the TM-doped black phosphorenes is very similar to that for the TM-doped blue phosphorenes [see Fig. 2(f)].

\begin{table*}
\caption{Spin moments in the TM impurity ($S_M$) and the nearest phosphorus neighbors ($S_{P1}$ and $S_{P2}$) for different substitutional TMs in black and blue phosphorenes, together with the spin moments of the isolated TM atoms ($S_{iso-atom}$). $S_{tot}$ is the total spin moment of the doped black and blue phosphorenes. The band gaps ($E_{g}$) of TM-doped black (blue) phosphorenes are also given. The values in parentheses are the PBE + \emph{U} band gaps.}
\begin{tabular}{p{1.2cm}p{2.0cm}p{1.5cm}p{1.5cm}p{1.5cm}p{1.5cm}p{2.5cm}p{2cm}}
\hline
\hline
      &Doped-atom  &   $S_M$ ($\mu_B$)   & $S_{P1}$ ($\mu_B$)  &  $S_{P2}$ ($\mu_B$)   &  $S_{tot}$ ($\mu_B$)   &   $S_{iso-atom}$ ($\mu_B$) &   $E_{g}$ (eV)  \\
\hline
        &Sc          &         &          &          &        &        &  0.97 (1.36)\\
        &Ti          &  0.986  &  -0.014  &  -0.017  &  1.00  &  3.00  &  0.36 (0.90)\\
        &V           &  1.977  &  -0.001  &  -0.002  &  2.00  &  5.00  &  0.07 (0.09)\\
        &Cr          &  3.082  &  -0.076  &  -0.085  &  3.00  &  6.00  &  0.72 (0.93)\\
 black-P&Mn          &  2.207  &  -0.053  &  -0.060  &  2.00  &  5.00  &  0.39 (0.82)\\
        &Fe          &  1.097  &  -0.036  &  -0.019  &  1.00  &  4.00  &  0.27 (0.90)\\
        &Co          &         &          &          &        &        &  0.61 (1.08)\\
        &Ni          &  0.953  &   0.064  &  -0.007  &  1.00  &  1.00  &  0.09 (0.46)\\
\hline
        &Sc          &         &          &          &        &     &  1.35 (1.59)\\
        &Ti          &  0.992  &  -0.020  &  -0.020  &  1.00  &  \  &  0    (0.73)\\
        &V           &  2.032  &  -0.055  &  -0.055  &  2.00  &  \  &  0.15 (0.47)\\
        &Cr          &  3.147  &  -0.083  &  -0.083  &  3.00  &  \  &  0.91 (1.63)\\
 blue-P &Mn          &  1.954  &  -0.032  &  -0.020  &  2.00  &  \  &  0.12 (0.73)\\
        &Fe          &  1.247  &  -0.044  &  -0.044  &  1.00  &  \  &  0.35 (0.91)\\
        &Co          &         &          &          &        &     &  0.69 (1.12)\\
        &Ni          &  0.967  &   0.039  &   0.039  &  1.00  &  \  &  0    (0) \\
\hline
\hline
\end{tabular}
\end{table*}

\textbf{3.2 Spin moments}

The spin moments of substitutional TMs in black and blue phosphorenes are displayed in Fig. 3(a), together with those of the isolated TM atoms. We find that the spin moments of the isolated TM atoms are 1, 3, 5, 6, 5, 4, 3, and 1 $\mu_B$ from Sc to Ni, respectively. On the other hand, the TM-substituted black phosphorenes have the zero magnetic moment for Sc and Co, but 1, 2, 3, 2, 1, and 1 $\mu_B$ for Ti, V, Cr, Mn, Fe, and Ni, respectively, which are the same as the corresponding cases in blue phosphorene. For both of the TM-substituted black and blue phosphorenes, we analyze the charge transfer using Bader charge (see Table I). We find that for both cases, the TM atoms lose electron charges while the nearest phosphorous atoms gain electron charges. It is notable that the magnitudes of gained and lost charges decrease as the atomic number increases.

Interestingly, as shown in Fig. 3(a), the total spin moments have the integer values of 0, 1, 2, 3, 2, 1, 0, and 1 $\mu_B$ for Sc-, Ti-, V-, Cr-, Mn-, Fe-, Co-, and Ni-doped black and blue  phosphorenes, respectively.

According to a recent first-principles study of substitutional TM impurities in graphene,\cite{Santos} the spin moments are calculated to be 0, 0, 1, 2, 3, 2, 1, and 1 $\mu_B$ for Sc-, Ti-, V-, Cr-, Mn-, Fe-, Co-, and Ni-doped graphene systems, respectively. These values are well compared with 0, 1, 2, 3, 2, 1, 0, and 1 $\mu_B$ for Sc-, Ti-, V-, Cr-, Mn-, Fe-, Co-, and Ni-doped black and blue phosphorenes, respectively. It is interesting to note that the spin moment of each TM impurity (except Sc and Ni) in graphene is smaller by 1 $\mu_B$ compared to the corresponding one in black and blue phosphorenes. This may be attributed to the different bonding natures of graphene and phosphorene: i.e., \emph{sp}$^{2}$ bonding in graphene and \emph{sp}$^{3}$ bonding in phosphorene. Since one valence electron of TM impurities in graphene participates in $\pi$ bonding with neighboring C atoms, the spin moment is likely to decrease by 1  $\mu_B$. To understand this deeply, we draw the schematic diagram of spin moment according to Hund's rule in Fig. 3(b). We here note that the valence electron configurations of Sc, Ti, V, Cr, Mn, Fe, Co and Ni are 3$d^1$4$s^2$, 3$d^2$4$s^2$, 3$d^3$4$s^2$, 3$d^5$4$s^1$, 3$d^5$4$s^2$, 3$d^6$4$s^2$, 3$d^7$4$s^2$, 3$d^8$4$s^2$, respectively. Briefly, we can distinguish the several regimes depending on the filling of electronic levels:

(i) Sc-doped black phosphorene have the empty Sc-phosphorous bonding levels, leading to a zero spin moment.

(ii) Co-doped black phosphorene have fully occupied Co-phosphorous bonding levels, leading to a zero spin moment.

(iii) Ti-, V-, and Cr-doped black phosphorene have partially occupied TM-phosphorous bonding levels with the spin moments of 1.00, 2.00, and 3.00 $\mu_B$, respectively.

(iv) Mn-, Fe-, and Ni-doped black phosphorene have partially occupied non-bonding 3\emph{d} levels with the spin moments of 2.00, 1.00, and 1.00 $\mu_B$.

It is notable that the spin moments of TM-doped blue phosphorene [see Fig. 3(c)] are the same as those of black phosphorene because the energy states of \emph{s} and \emph{d} in the outermost orbital of TM atoms and phosphorus atom are very close to each other.

\begin{figure}[htb]
\includegraphics[width=6.5cm]{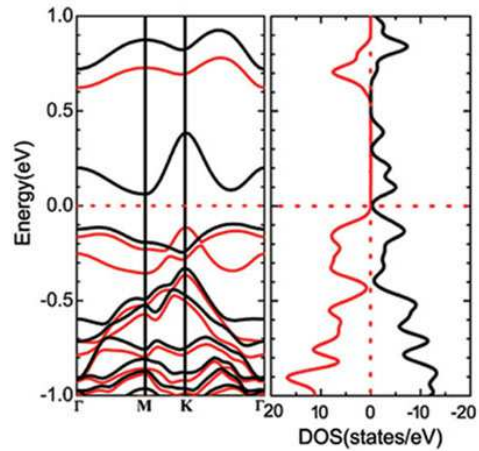}
\caption{ (color online). Band structure and density of states of the undoped defective black phosphorene. The red lines represent majority spin band, while the black lines represent minority spin band. The energy zero represents the Fermi level.}
\end{figure}

To explore the underlying mechanism of the magnetic moments in TM-doped black and blue phosphorenes, the Mulliken population analysis was performed to list the results in Table II. We find that the spin moments of the TM impurities ($S_M$) have a dominant contribution for the nearest phosphorus neighbors ($S_{P1}$ and $S_{P2}$). The calculated spin moments of TM impurities for Ti, V, Cr, Mn, Fe and Ni in doped black phosphorene are $S_M$ = 0.986, 1.977, 3.082, 2.207, 1.097, and 0.953 $\mu_B$, respectively, close to the above-discussed Hund's analysis. Similarly, the spin moments of TM impurities for Ti, V, Cr, Mn, Fe and Ni in doped blue phosphorenes are $S_M$ = 0.992, 2.032, 3.147, 1.954, 1.247 and, 0.967 $\mu_B$, respectively.

\begin{figure*}[htb]
\includegraphics[width=16cm]{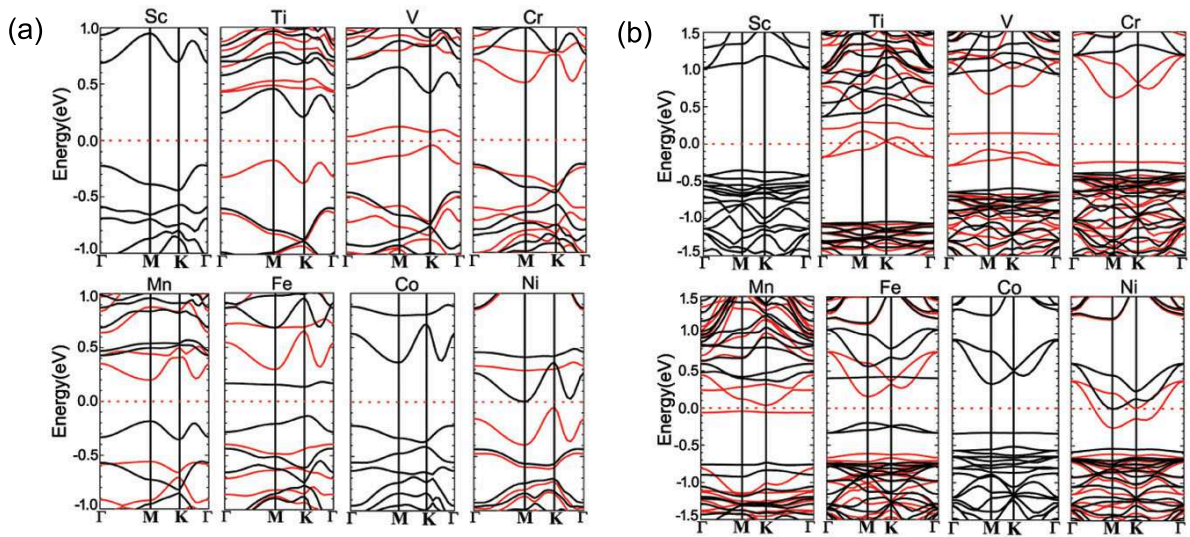}
\caption{ (color online). Band structures of Sc-, Ti-, V-, Cr-, Mn-, Fe-, Co- and Ni-doped black (a) and blue (b) phosphorenes, respectively. The red (black) lines represent the majority (minority) spin band. The energy zero represents the Fermi level.}
\end{figure*}

\begin{figure}[htb]
\includegraphics[width=7cm]{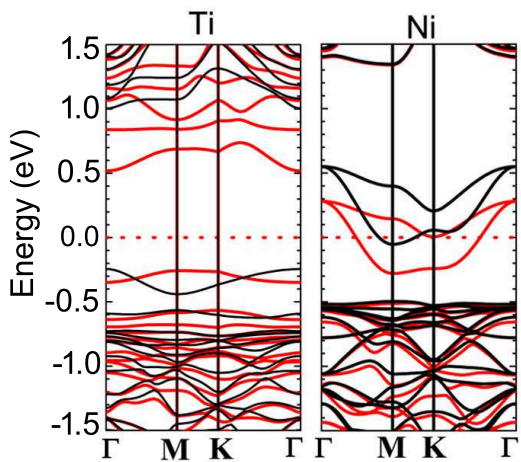}
\caption{ (color online). Band structures of Ti-, Ni-doped blue phosphorenes obtained using the PBE + \emph{U} calculation with \emph{U} = 5.5, 6.5 eV, respectively. The red (black) lines represent the majority (minority) spin band. The energy zero represents the Fermi level.}
\end{figure}

\textbf{4. Analysis of the electronic structures}

We first examine the electronic structure of a single phosphorous vacancy in black phosphorene. As substitutional impurities in black phosphorene, most of the TM atoms studied here exhibit a symmetrical configuration of C$_{2h}^{3}$. \cite{Ribeiro} For this reason, it is particularly instructive to analyze their electronic structures with the hybridization between the atomic levels of the TM atoms and those associated with a relaxed C$_{2h}^{3}$ symmetrical phosphorus vacancy. As shown in Fig. 4, the C$_{2h}^{3}$ vacancy shows a considerable spin polarization of 1.00 $\mu_B$, indicating a dilute magnetic property.

\begin{figure*}[htb]
\includegraphics[width=16cm]{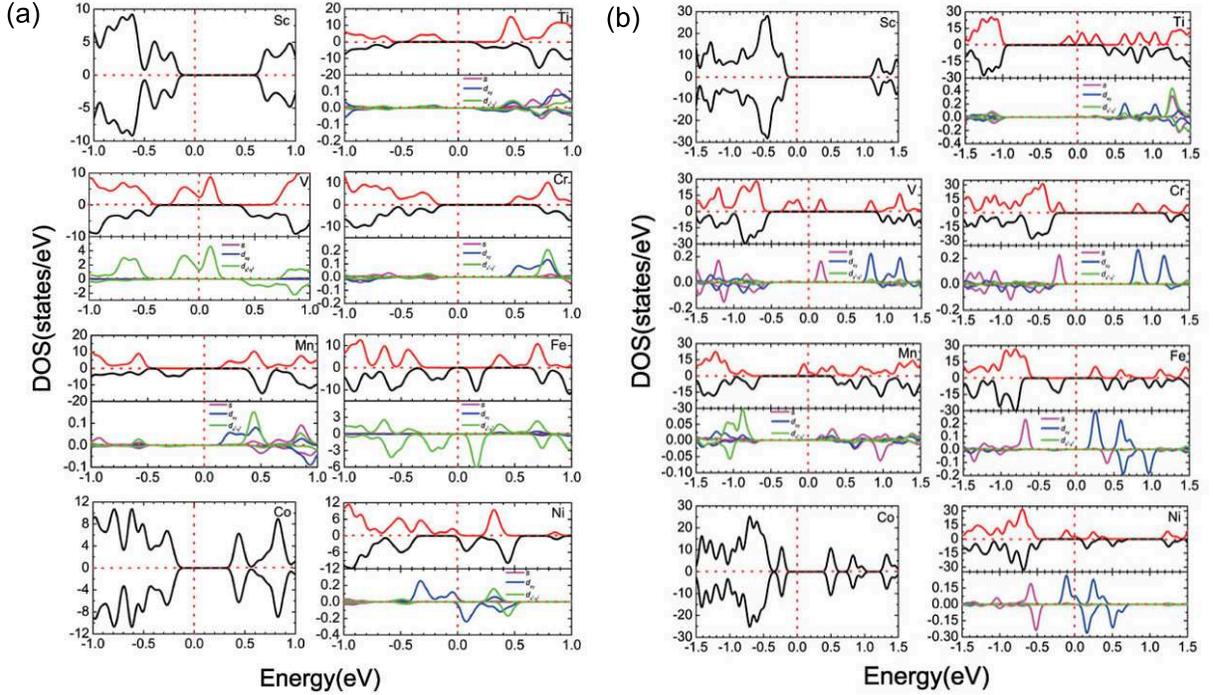}
\caption{(color online). Spin polarized total (upper panel) and partial (lower panel) density of states of Sc-, Ti-, V-, Cr-, Mn-, Fe-, Co- and Ni-doped black (a) and blue (b) phosphorenes, respectively. The energy zero represents the Fermi level.}
\end{figure*}

To further shed light on the underlying mechanism of magnetic properties in the TM-doped black and blue phosphorene structures, we plot the spin-polarized band structures of TM-doped black and blue phosphorenes in Fig. 5(a) and (b), respectively. Interestingly, the majority and minority spin bands for Ti-, V-, Cr-, Mn-, Fe-, and Ni-doped black phosphorene show semiconducting characters [see Fig. 5(a)], indicating dilute magnetic properties. On the other hand, Sc- and Co-doped black phosphorene exhibit zero spin moment, whereas Ti- and Ni-doped blue phosphorene show half-metal characters [see Fig. 5(b)]. Note that V-, Cr-, Mn-, and Fe-doped blue phosphorenes exhibit dilute semiconducting characters, while Sc- and Co-doped blue phosphorenes have zero spin moment.

In general, substitutional TM impurities in black and blue phosphorenes exhibit very similar behaviors in their energetic and magnetic properties. This result indicates that the structural differences of black- and blue-phosphorene lattices are insensitive to determine the energetic and magnetic properties of TM-doped black and blue phosphorenes, as shown in Fig. 2, 3, and 5.

It is interesting to examine the effect of the on-site Coulomb interaction \emph{U} on the magnetic properties of the substitutional 3\emph{d} TM impurities in black and blue phosphorenes. We perform the PBE + \emph{U} calculations for all the considered systems, where the values of \emph{U} = 4.0, 5.5, 3.3, 3.5, 3.5, 4.3, 3.3, and 6.5 eV are chosen for Sc-, Ti-, V-, Cr-, Mn-, Fe-, Co-, and Ni-doped systems, respectively.\cite{Rai,Huang,Pari} The calculated PBE + \emph{U} band gaps (\emph{Eg}) of Sc-, Ti-, V-, Cr-, Mn-, Fe-, Co-, and Ni-doped black and blue phosphorenes are listed in Table II. We find that \emph{Eg} of the magnetic semiconductor obtained using PBE + \emph{U} increases by $\sim$30\% compared to the PBE results. However, it is noticeable that the spin moment does not change depending on the PBE and PBE + \emph{U} methods. Interestingly, we find that the PBE + \emph{U} band structure of Ti-doped blue phosphorene shows a magnetic semiconductor property with a gap opening (see Fig. 6), different from the half-metallic character predicted by PBE. This indicates that \emph{U} in Ti-doped blue phosphorene splits the narrow half-filled bands crossing the Fermi level into lower and upper Hubbard bands. On the other hand, the half-metallic character of Ni-doped blue phosphorene predicted by PBE is preserved in the PBE + \emph{U} band structure (see Fig. 6), because the bands crossing the Fermi level have relatively larger band widths compared to those in Ti-doped blue phosphorene [see Fig. 5(b)].

A general picture of the dilute magnetic and half-metal features of the TM-doped black and blue phosphorenes can be seen from the analysis of the spin-polarized total and partial DOS, as shown in Fig. 7. As for Sc- and Co-doped black phosphorenes, the total DOS of the majority and minority states are completely compensated with each other, yielding zero spin moment [see Fig. 7(a)]. It is found that for Ti-, V-, Cr-, Mn-, Fe- and Ni-doped black phosphorenes, the total DOS of the majority and minority states are not compensated below $E_F$ and show a gap opening, indicating dilute magnetic semiconducting properties. We note that the DOS of Sc- and Co-doped blue phosphorene show nonmagnetic properties; those of Ti- and Ni-doped blue phosphorene show half-metal properties; V-, Cr-, Mn-, Fe-doped blue phosphorenes show dilute magnetic semiconductor characters [see Fig. 7(b)]. From the analysis of the spin-polarized total and partial DOS, it is seen that the magnetic moments are well localized at the TM atom site, and the $d_{xy}$ and $d_{x^2-y^2}$ orbitals are dominant for the contribution to the partial DOS.

\textbf{5. Conclusions }

We have performed a first-principles DFT calculation for the structural, energetic, and magnetic properties of a series of substitutional 3\emph{d} TM impurities in black and blue phosphorenes. We provided a simple model based on Hund's rule for understanding the calculated electronic and magnetic properties of the considered systems, where the dilute-semiconducting and half-metal features, spin moment, and binding energy are varied depending on the atomic number of the TM atoms. The spin-polarized band structures and DOS calculations show that for black phosphorene, the Ti-, V-, Cr-, Mn-, Fe- and Ni-doped systems have dilute magnetic semiconductor properties, while Sc- and Co-doped systems have no magnetism. For blue phosphorene, the Ti- and Ni-doped systems show half-metal properties, while V-, Cr-, Mn- and Fe-doped systems show dilute magnetic semiconductor characters, Sc- and Co-doped systems show non-magnetism.

Since substitutional impurities of 3\emph{d}  TM atoms in black and blue phosphorenes exhibit some intriguing electronic and magnetic properties, such doped systems can provide an interesting route to tune various functions for spin electronic devices based on black and blue phosphorenes. This functional ability together with the high stability of substitutional impurities can open a route to fabricate ordered arrays of these impurities at predefined locations, which would allow the experimental tests of the theoretical predictions of unusual magnetic interactions mediated by black and blue phosphorenes.

\textbf{Acknowledgements}

We thank Prof. Zhenyu Zhang for helpful discussions. This work was supported by the National Basic Research Program of China (Grant No. 2012CB921300), National Natural Science Foundation of China (Grant Nos. 11274280 and 11304288), and National Research Foundation of Korea (Grant No. 2015R1A2A2A01003248).

\end{document}